\documentclass[aps,groupedaddress,showpacs,11pt,nofootinbib]{revtex4}
\usepackage[dvips]{graphicx}
\usepackage{amsmath}
\usepackage{amssymb}
\usepackage[dvips]{graphicx}
\usepackage[]{caption}
\usepackage{amsmath}
\usepackage{amssymb}
\usepackage{dcolumn}
\usepackage{bm}
\usepackage{amsfonts}
\usepackage{amsthm}
\usepackage{natbib}
\usepackage{fix2col}

\newcommand{\bea}{\begin{eqnarray}}
\newcommand{\eea}{\end{eqnarray}}
\newcommand{\be}{\begin{equation}}
\newcommand{\ee}{\end{equation}}

\newcommand{\half}{\frac{1}{2}}
\newcommand{\bi}{\begin{itemize}}
\newcommand{\ei}{\end{itemize}}

\begin{document}
\bibliographystyle{prsty}
\title{
    {\bf\Large
        Modulated Perturbations from Instant Preheating after new Ekpyrosis 
    }
    \vspace{1cm}
}
 \vspace{1cm}
\author{Thorsten Battefeld}
\email{tbattefe@princeton.edu}
\affiliation{ Princeton University,
Department of Physics,
NJ 08544
}

\begin{abstract}

\vspace{1cm}

\noindent
We present a mechanism to transfer the spectrum of perturbations in a scalar isocurvature field $\xi$ onto the matter content in the radiation era via modulated, instant preheating after ekpyrosis. In this setup, $\xi$ determines the coupling constant relevant for the decay of a preheat matter field into Fermions. The resulting power spectrum is scale invariant if $\xi$ remains close to a scaling solution in new ekpyrotic models of the universe; by construction the spectrum is independent of the detailed physics near the bounce. The process differs from the curvaton mechanism, which has been used recently to revive the ekpyrotic scenario, in that no peculiar behavior of $\xi$ shortly before or during the bounce is needed. In addition,  a concrete and efficient realization of reheating after ekpyrosis is provided; this mechanism is not tied to ekpyrotic models, but could equally well be used in other setups, for instance inflationary ones. We estimate non-Gaussianities and find no additional contributions in the most simple realizations, in contrast to models using the curvaton mechanism.

\end{abstract}
\maketitle

\newpage
\section{Introduction}

The ekpyrotic scenario \cite{Khoury:2001wf,Khoury:2001bz,Steinhardt:2001st} is an attempt to provide an alternative to the inflationary model of the early universe, see \cite{Linde:2007fr} for a review. It is based on the moduli space approximation of heterotic M-theory in five dimensions. In the proposal, the fifth dimension is sandwiched between two orbifold fixed planes and undergoes a bounce, corresponding to a bounce of the scale factor in the four dimensional effective description as well. This can provide an alternative solution to the horizon problem \footnote{See \cite{Buchbinder:2007tw} for an updated account on the proposed resolution of the horizon and flatness problem within ekpyrotic proposals, attempting to address objections raised in e.g. \cite{Linde:2007fr}.}. However, the model is restrained by the singular nature of the bounce, which renders the resulting scalar spectral index for the radion ambiguous \cite{Brandenberger:2001bs,Durrer:2002jn,Creminelli:2004jg} \footnote{For instance, if the matching of perturbations is done on a constant energy hypersurface, a deep blue spectrum results \cite{Brandenberger:2001bs,Deruelle:1995kd}.}, even from the five dimensional point of view \cite{Battefeld:2004mn}.

Recently, the model has been  revived
\cite{Lehners:2007ac,Buchbinder:2007ad,Creminelli:2007aq,Buchbinder:2007tw} by
considering additional degrees of freedom, such as the volume modulus
of the Calabi-Yau. If the fields start out near an (unstable) scaling solution, nearly scale invariant fluctuations are
generated in an isocurvature mode during the pre-bounce phase. These can be converted to adiabatic ones before the bounce by different means; for example, the fields could be moving away from the multi field scaling solution towards a single field ekpyrotic attractor shortly before the bounce \cite{Koyama:2007ag,Koyama:2007mg,Koyama:2007if}, naturally requiring fine-tuning (see however \cite{Buchbinder:2007tw}). Alternatively, a reflection of fields on sharp boundaries,  which can be argued to occur  naturally in certain models \cite{Lehners:2006pu,Lehners:2006ir,Lehners:2007nb,Lehners}, can cause a similar conversion.  The
idea of employing isocurvature perturbations resembles the curvaton mechanism \cite{Lyth:2001nq},
and has already been proposed  by Notari and Riotto in
\cite{Notari:2002yc} shortly after problems with the ekpyrotic
scenario surfaced.

However, there remains a subtlety regarding the bounce itself: if the low
energy effective theory in four dimensions stays valid throughout,
energy conditions must necessarily be violated in order for a
non-singular bounce to occur. The only known mechanism capable of achieving this without introducing fatal instabilities is by means of a ghost condensate \cite{Buchbinder:2007ad,Creminelli:2007aq,Buchbinder:2007tw}, whose connection to the underlying string theory is unknown. Further, since the conversion of isocurvature to adiabatic perturbations needs to occur towards the end of ekpyrosis, both degrees of freedom need to evolve quite drastically near the not well understood bounce of the four dimensional scale factor \footnote{Note that the bounce mentioned in \cite{Lehners:2007ac} is this sudden change of the trajectory in field space and not the bounce of the four dimensional scale factor which is needed to solve the horizon problem. The latter bounce is still assumed to be singular in \cite{Lehners:2007ac}.}. Hence  one might be concerned that the nature of the bounce has an impact on the spectrum of scalar perturbations, an issue studied extensively in four dimensional toy models, see e.g. \cite{Finelli:2007tr,Bozza:2005xs,Bozza:2005wn,Geshnizjani:2005hc,Battefeld:2005cj} and references therein.

In this article, we propose a mechanism to imprint the spectrum of isocurvature perturbations in a field $\xi$ onto the matter content during reheating, that
is during the conversion of some preheat matter content originating at
the brane collision into relativistic Fermions. This provides the possibility to generate scale invariant adiabatic perturbations after the not well understood bounce. The conversion is achieved by assuming a
sensitivity of the coupling constant between the Fermions and a
preheat matter field on  the value of $\xi$. This determination of
coupling constants in terms of more fundamental fields is a natural feature of
string theory. We take the isocurvature field to be fixed throughout the bounce and pre-bounce phase in order to
remain faithful to the original ekpyrotic proposal. In other words, to touch base with the proposals in
\cite{Lehners:2007ac,Buchbinder:2007ad,Creminelli:2007aq,Koyama:2007ag,Koyama:2007mg}, we stay close to the multi-field scaling solution.
In order to arrive at a scale invariant spectrum for $\xi$, a close proximity to this ill-fated unstable solution in the ekpyrotic phase is necessary. As a prerequisite, either initial conditions have to be extremely fine-tuned, or an additional pre-ekpyrotic phase needs to be added during which $\xi$ is driven towards the scaling solution \cite{Buchbinder:2007tw}. Our proposal is insensitive to details of the bounce by construction, and a conversion of isocurvature to adiabatic perturbations before the bounce is not needed.

  The employed mechanism of imprinting perturbation during reheating is known as modulated preheating in the inflationary literature \cite{Kofman:2003nx,Dvali:2003em,Bernardeau:2004zz} (see also \cite{Enqvist:2003uk,Tsujikawa:2003bn,Matarrese:2003tk,Mazumdar:2003iy,Postma:2003jd,Allahverdi:2004bk,Ackerman:2004kw,Vernizzi:2005fx,Lyth:2005qk}), but it has not yet been applied to ekpyrotic scenarios (even though briefly mentioned in \cite{Tsujikawa:2003bn}) or instant preheating, which we employ.

In the ekpyrotic proposal, preheating is due to the collision
of boundary branes with two salient features: first, the event is
nearly instantaneous, since it is directly related to the collision of
branes; once they come close to each other, new light states appear and
get produced non-perturbatively. Second, the force between the
boundary branes vanishes near the collision
\cite{Steinhardt:2001st}.

 To model these features from a four
dimensional point of view, we use the scenario of instant preheating
\cite{Felder:1999pv} where a scalar preheat matter field $\chi$,
which can be thought of as one of the additional light degrees of freedom that appear near the bounce, gets produced once the adiabatic field $\phi$ is close to zero \footnote{Note that $\phi$ is decreasing in the contracting phase; this ekpyrotic phase ends around $\phi\sim 0$, the field runs quickly to $-\infty$, turns around (corresponding to the bounce), and emerges with $\dot{\phi}>0$ in the post-bounce phase.}. As $\phi$ evolves towards larger values again, $\chi$-particles become heavy and decay into relativistic matter, which we take to be a Fermionic field $\psi$. The time of $\chi$-decay will depend on the value of the isocurvature field if we introduce a dependence of the coupling constant on $\xi$. This sensitivity leads to modulated perturbations \cite{Kofman:2003nx,Dvali:2003em,Bernardeau:2004zz}.

The dependency of the coupling constant between
preheat-matter and Fermions on the isocurvature field could in principle be computed if the
stringy construction, as well as the relation to the standard
model of particle physics, were well understood. Unfortunately, this is not the
case here and this contingency is kept arbitrary. As it turns out, this
dependence is not overly constrained from a model builders' perspective, as long as it is steep enough to yield the dominant contribution to the power spectrum.

To discriminate the above model from \cite{Lehners:2007ac,Buchbinder:2007ad,Creminelli:2007aq}, we compute non-Gaussianities (NG); we find no additional contributions due to the conversion in the simplest case of an exponential dependence, in contrast to set-ups involving the curvaton mechanism (see e.g. \cite{Sasaki:2006kq} and references therein) or the recently proposed new ekpyrotic scenarios \cite{Creminelli:2007aq,Buchbinder:2007tw,Koyama:2007mg} \footnote{If the conversion is due to a reflection of fields on sharp boundaries, the resulting NG can be much smaller too \cite{Lehners}.}. However, larger Non-Gaussianities can occur if the dependence of the coupling on $\xi$ is more involved.

\section{Modulated Perturbations from Instant Preheating after Ekpyrosis}
In single field ekpyrotic scenarios, the radion describes the inter-brane distance from a four dimensional point of view. However, other degrees of freedom such as the Calabi-Yau volume modulus are present and also expected to be dynamic during the contracting phase (corresponding to a decreasing radion). In such a multi-field solution, one can always identify a single degree of freedom along the trajectory in field space, the adiabatic mode $\phi$, and fields perpendicular to the trajectory, the isocurvature modes. For simplicity, we consider two fields only and denote the isocurvature field with $\xi$, see Appendix \ref{ap2} for details. Given a scaling solution in the pre-bounce phase as in \cite{Lehners:2007ac,Buchbinder:2007ad,Creminelli:2007aq,Koyama:2007ag,Koyama:2007mg}, the isocurvature field acquires a nearly scale invariant spectrum of fluctuations, as reviewed in Appendix \ref{ap1}. If the scaling solution is abandoned before the bounce and the old, single field ekpyrotic attractor is approached, a conversion of isocurvature perturbations to adiabatic ones takes place simultaneously. As a consequence, large non-Gaussianities are produced \cite{Koyama:2007mg}. To avoid large non-Gaussianities we consider the absence of any conversion before or during the bounce. This is only the case if the trajectory in field space is not curved, that is if the fields remain close enough to the (unstable) scaling solution throughout \cite{Koyama:2007ag,Koyama:2007mg}. Thus, we take $\xi$ as approximately constant by construction \footnote{We make this assumption primarily to focus on the effects due to modulated instant preheating, not because it is needed for the proposed mechanism to work; one could conceivably abandon the scaling solution before the bounce and still imprint perturbations during preheating as discussed in this article.}, not because it is stabilized; indeed, a tachyonic direction is unavoidable in the new ekpyrotic scenario, if the isocurvature mode is to have a nearly scale invariant spectrum of fluctuations \cite{Tolley:2007nq}, see also Appendix \ref{ap2}.

Consider now the approach of the boundary branes: we expect the force between them to vanish when they are close to each other \cite{Steinhardt:2001st}, so that the potential for the radion disappears.
Thus, we take $V(\phi,\xi)\rightarrow 0$ for $\phi\rightarrow 0$, in accord with ekpyrotic proposals \cite{Steinhardt:2001st}. After the collision ($\phi\rightarrow -\infty$ and back again) the universe emerges with an increasing $\phi$.

Given this generic evolution of ekpyrotic models, we
would like to focus on the preheating stage occurring when $\phi\sim
0$, that is, when branes are within a string length of each
other (See \cite{Traschen:1990sw,Shtanov:1994ce,Kofman:1994rk,Kofman:1997yn} as well as the review \cite{Bassett:2005xm} and references therein for a sample of the extensive literature on (p)reheating). Being in such close proximity, new light degrees of freedom are expected to be present \cite{Kofman:2004yc}. For simplicity, to model these new degrees of freedom near $\phi\sim 0$, we consider a single bosonic pre-heat matter field $\chi$ coupled to the adiabatic field $\phi$. Additionally,  to achieve reheating of ordinary matter, we couple $\chi$ to a fermionic field $\psi$. Thus, we take the effective potential
\begin{eqnarray}
V_{\text{eff}}=\half g^2 \phi^2 \chi^2+h \chi \psi \bar{\psi}\,,
\end{eqnarray}
to describe preheating and reheating, for $\phi \geq 0$ and increasing. Assuming an appropriate value of $g$, the adiabatic field can produce $\chi$-particles when it is close to zero. This production can occur rather quickly via non-perturbative effects in a small region around the origin \cite{Felder:1999pv}, say for $\phi\leq \phi_0$. After particle production ($\phi>\phi_0$) the $\chi$ particles become non-relativistic. Further, since the potential vanishes for $\phi\sim 0$, the adiabatic field is effectively rolling freely in a universe dominated by kinetic energy. Here, we made the conservative assumption that only a small fraction of the energy in the adiabatic field gets transferred to $\chi$ particles initially. Since the equation of state parameter for such a stiff fluid is $w=1$ we have $a\sim t^{1/3}$. With this time dependence in mind, we can  integrate the Klein-Gordon equation for $\phi$ to
\begin{eqnarray}
\phi=\frac{M_p}{2\sqrt{3\pi}}\ln(t/t_0)+\phi_0\,,
\end{eqnarray}
where $t_0=1/3H_0\approx M_p/2\sqrt{3\pi}\dot{\phi}_0$ is the Hubble time around particle creation, depending on the initial velocity $\dot{\phi}_0 >0$ of the adiabatic field just before the production of $\chi$-particles. The energy in the adiabatic field redshifts very fast $\propto a^{-6}$ thereafter. Since $\phi_0\ll M_P$, we are justified to work in the 'instantaneous' limit, that is with $\phi_0\approx 0$; this corresponds to the fast creation of particles once the branes are within a string length of each other. We can also neglect backreaction of $\chi$ particles, which would become important at $t_1$ when $\phi\sim M_P$, since $\chi$ particles decay earlier for appropriate values of $h$ at  $t_c<t_1$. Up until $t_c$ the energy in the preheat matter field is increasing, due to the increase of $\phi$ to which it is coupled. If no decay channel were open to $\chi$ the adiabatic field would bounce back due to backreaction, providing an example of moduli trapping \cite{Kofman:2004yc,Watson:2004aq,Patil:2004zp,Battefeld:2005wv,Battefeld:2005av,Battefeld:2006cn,Cremonini:2006sx,Greene:2007sa}. This is not the case here, since $\chi$-particles can decay into Fermions. Thus, the time $t_c$ marks the transition to a radiation dominated phase. In the cyclic scenario, the interbrane potential for the radion would kick in at some point later, slowing it down until it eventually turns around in the far future to commence another cycle \cite{Steinhardt:2001st,Erickson:2006wc}. However, it is not clear to us how a cyclic evolution occurs in the presence of more fields, especially in the presence of an unstable direction. Accordingly, we will be content with a single bounce in this study.

The preheating stage described above is the instant preheating scenario of \cite{Felder:1999pv} proposed within the inflationary framework; we refer the interested reader to \cite{Felder:1999pv} for more details, e.g. regarding constraints on coupling constants $g$ and $h$ such that  sufficient production of $\chi$ and $\psi$ particles takes place. The new feature above is the implementation of instant preheating within the ekpyrotic framework.

Next, consider a dependence of the coupling "constant" $h$ on the isocurvature  field $\xi$, which has been a spectator during the bounce and  the preceding ekpyrotic phase \footnote{In principle $g$ could also depend on $\xi$, but since $g$ is heavily constraint from requiring that $\chi$-particles get produced sufficiently, we will ignore such a dependence.}. Such a dependence is indeed expected in many stringy models, but a specific construction is lacking so far. Taking this uncertainty as an opportunity for optimism, we regard $h(\xi)$ as a free function, constraint by the requirement that $h(\xi_{scaling})$ yields sufficient production of Fermions. That way we can provide a new realization of modulated perturbations from reheating \cite{Kofman:2003nx,Dvali:2003em,Bernardeau:2004zz}; the decay rate of $\chi$-particles depends on this coupling and the radion field value via $\Gamma\propto h^2 \phi$ \cite{Felder:1999pv}. Since $\phi$ is only increasing logarithmically, the decay rate before $t_1$ is very well described by its value at $t_1$. Henceforth, the only dependence on the fermionic coupling is
\begin{eqnarray}
\Gamma\propto h^2\,,
\end{eqnarray}
so that the time at which Fermions are produced becomes
\begin{eqnarray}
t_c(\xi)\propto h(\xi)^{-2}\label{t_c}\,. \label{t_cofh}
\end{eqnarray}
The corresponding adiabatic field value at $t_c$ depends only logarithmically on $h$ via
\begin{eqnarray}
\phi_c=\frac{M_p}{2\sqrt{3\pi}}\ln(h^{-2})+\hdots\,,
\end{eqnarray}
where the dots denote terms independent of $h$.
The dependence on $h(\xi)$ in (\ref{t_cofh}) induces an additional contribution to the power spectrum and higher order correlation functions.

To compute these effects, we use the $\delta N$-formalism, first proposed by Starobinsky in \cite{Starobinski} and extended by Sasaki and Stewart \cite{Sasaki:1995aw} and others \cite{Nakamura:1996da,Lyth:2004gb,Lyth:2005fi,Seery:2005gb,Vernizzi:2006ve,Byrnes:2006vq,Byrnes:2007tm,Battefeld:2006sz,Battefeld:2007en}. This formalism relies on the fact that on super-Hubble scales perturbations evolve classically (no oscillations) and fluctuations can be taken smooth on scales smaller than the Hubble radius. Based on this, one can relate the perturbation of the volume expansion rate $\delta N$ to the curvature perturbation $\zeta$ (which is conserved on large scales in simple models) even beyond linear order \cite{Lyth:2003im,Rigopoulos:2003ak} \footnote{Note that the separate Universe formalism developed by Rigopoulos and Shellard in e.g. \cite{Rigopoulos:2003ak} is equivalent to the $\delta N$-formalism.}. Given this relationship, correlation functions of $\zeta_k$ ($k$ denotes a Fourier mode) can be computed in terms of the change of $N$ during the evolution of the Universe. The formalism cannot only be applied to simple models of inflation, but also to scenarios where fluctuations are imprinted on $\zeta$ after reheating, e.g. the curvaton scenario \cite{Lyth:2005fi}, and provides an extremely efficient way of estimating non-Gaussianities (here the non linear $\delta N$-formalism developed in \cite{Lyth:2004gb,Lyth:2005fi} is needed).

Following this thread of thought, the resulting power-spectrum for two fields becomes
\begin{eqnarray} \label{pwands}
{{P}}_{\zeta}=\left( \frac{\partial N}{\partial \phi}  \right)^2 {{P}}_\phi +  \left( \frac{\partial N}{\partial \xi}  \right)^2 {{P}}_{\xi}\,,
\end{eqnarray}
where the derivatives are to be taken deep inside the pre-bounce phase. Extending the $\delta N$-formalism to more fields and higher orders results in
\begin{eqnarray} \label{deltan}
\zeta(t_2, \vec{x})= \sum_I \frac{\partial N}{\partial \phi_I} \delta \phi_I+\frac{1}{2}\sum_{IJ} \frac{\partial N}{\partial \phi_I} \frac{\partial N}{\partial \phi_J} \delta \phi_I\delta \phi_J + \dots \,,
\end{eqnarray}
so that higher order correlation functions, such as the bi- or trispectrum, can be computed too (see e.g. \cite{Byrnes:2006vq,Byrnes:2007tm}).

Consider now the number of e-folds from the production of $\chi$-particles until after their decay into Fermions, when we are deep in the radiation era, specifically at $t_r$
\begin{eqnarray}
\nonumber N_r(h) &\equiv&\int_{t_0}^{t_r} H\, dt\\
\nonumber &=&\int_{t_0}^{t_c^-(h)} H_-\, dt+\int_{t_c^+(h)}^{t_r} H_+\, dt\\
&=&\int_{\eta_0}^{\eta_c^-(h)} \frac{r_-}{\eta}\, d\eta+\int_{\eta_c^+(h)}^{\eta_r} \frac{r_+}{\eta}\, d\eta\,. \label{dummy}
\end{eqnarray}
Here, we switched to conformal time $\eta$ and used the Hubble parameter $\mathcal{H}=a^\prime/a=r/\eta$ with $r_-=1/2$ (kinetic driven phase) or $r_+=1$ (radiation phase). Conformal time makes a jump at the transition, $\eta_c^+=\eta_c^-r_+/r_-=2\eta_c^-$, since $\mathcal{H}$ is continuous at the transition via the Israel junction conditions \cite{Israel:1966rt,Deruelle:1995kd}. Integrating (\ref{dummy}) yields
\begin{eqnarray}
N_r=r_-\ln(\eta_c^-/\eta_0)-r_+\ln(\eta_c^-r_+/\eta_rr_-)\,.
\end{eqnarray}
Taking the derivative with respect to $\xi$ and using $\eta_c^-\propto t_c^{-(r_-+1)}\propto h^{2(r_-+1)}$ we arrive at
\begin{eqnarray}
\nonumber \frac{\partial N_r}{\partial \xi}&=&(r_--r_+)\frac{1}{\eta_c^-}\frac{\partial \eta_c^-}{\partial h}\frac{\partial h}{\partial \xi}\\
&=&2(r_--r_+)(r_-+1)\frac{1}{h}\frac{\partial h}{\partial \xi}\,,\label{partialNpartialxi}
\end{eqnarray}
so that the additional contribution to the spectrum becomes
\begin{eqnarray}
P_{\zeta +}&=&\left( \frac{\partial N_r}{\partial \xi}  \right)^2 {{P}}_\xi \nonumber\\
\nonumber &=&\left(2(r_--r_+)(r_-+1)\frac{1}{h}\frac{\partial h}{\partial \xi}\right)^2{{P}}_\xi\,\\
&=&\left(\frac{3}{2}\frac{1}{h}\frac{\partial h}{\partial \xi}\right)^2{{P}}_\xi\,. \label{resultdN}
\end{eqnarray}
It should be noted that the more pedestrian method of using the Deruelle-Mukhanov matching conditions \cite{Israel:1966rt,Deruelle:1995kd} across a sharp transition of the equations of state parameter could be used alternatively \cite{unpublished}, but the $\delta N$-formalism is more efficient. The contribution in (\ref{resultdN}) can dominate over the one of the adiabatic field, if there is a steep dependence of $h$ on the primordial value of the isocurvature field $\xi$. Therefore, the relevant spectrum for the cosmic-microwave background radiation does not need to be the one of the adiabatic field, but can be provided by an isocurvature field (see Appendix \ref{ap2} for the computation of its spectral index) \footnote{During completion of this work, \cite{Suyama:2006rk} appeared providing an application of the $\delta N$-formalism to modulated perturbations in ordinary preheating after inflation -- see note added after the conclusions.}.

The difference to the curvaton scenario \cite{Lyth:2001nq} or the recently proposed new-ekpyrotic scenario \cite{Lehners:2007ac,Buchbinder:2007ad,Creminelli:2007aq} is simple; namely, to employ the curvaton mechanism one needs the additional degree of freedom to be dynamic in a very peculiar way close to the end of ekpyrosis/inflation, with the danger of being sensitive to the bounce in ekpyrotic models. In our scenario, $\xi$ remains constant throughout ekpyrosis, causing the decay of the pre-heat field $\chi$ to occur at slightly different times, in accordance with $\xi$'s (isocurvature) fluctuations. Thus, the proposed mechanism is rather robust and insensitive to the nature of the bounce by construction.

The main drawback of the ekpyrotic mechanism to refurbish the isocurvature field $\xi$ with a nearly scale invariant spectrum of fluctuations in the contracting phase is the necessity of $\xi$ to have a tachyonic direction \cite{Tolley:2007nq}.  As a result, $\xi$ is not stabilized, but it is supposed to linger near the crest of the potential. Even if our interest lies in a single bounce only, to be close to the scaling solution \cite{Koyama:2007ag,Koyama:2007mg,Tolley:2007nq} there is a need for highly fine-tuned initial conditions. As long as no compelling mechanism is found to explain such initial values, the model remains rather unpleasant, proving, once again, the difficulty of constructing alternative mechanisms to inflation with regards to the observed spectrum of fluctuations. One can by all means add a pre-ekpyrotic phase and postulate a potential such that $\xi$ is driven very close to $\xi_{scaling}$ \cite{Buchbinder:2007tw}, but it remains to be seen if such a potential arises in a concrete stringy construction.

We would like to emphasize that modulated perturbations from instant preheating as described above cannot only occur after an ekpyrotic phase, but they can also be present in an inflationary framework; if an isocurvature mode has a scale invariant spectrum of fluctuations after inflation (which is the case if additional light degrees of freedom are present during inflation), the desired scale invariance of $\zeta$ after reheating can be achieved without relying on the curvaton mechanism. This is only a slight extension of \cite{Kofman:2003nx,Dvali:2003em} \footnote{See also \cite{Byrnes:2005th} for a related model, where the decay rate depends on the phase of a complex scalar inflaton field.}.
Meanwhile one may naturally wonder how the above mechanism, which converts isocurvature perturbations to adiabatic ones, could be differentiated from a conversion before the bounce or from simple models of inflation. A possibility is to consider non-Gaussianities, namely, the non-linearity parameter $f_{NL}$ characterizing the ratio of the bispectrum divided by the square of the power spectrum. The observational bound on the full non-linearity parameter
obtained from the WMAP3 data set alone is $-54<f_{NL}<114$ \cite{Spergel:2006hy,
Creminelli:2005hu,Creminelli:2006rz}, and future experiments will
improve upon it considerably \cite{Hikage:2006fe,Komatsu:2001rj,Planck}. The momentum independent contribution to $f_{NL}$  can be computed within the $\delta N$-formalism to \cite{Vernizzi:2006ve}
\begin{eqnarray}
\frac{6}{5}f_{NL}=\frac{\sum_{I,J} N_{I}N_{J}N_{IJ}}{\left(\sum_{K} N^2_{K}\right)^2}\,, \label{fnl}
\end{eqnarray}
where the sum runs over all fields, $\phi$ and $\xi$ in our case, and $N_\xi=\partial N/ \partial \xi$, $\dots$. Non-Gaussianities from modulated perturbations after inflation have been considered in the past \cite{Zaldarriaga:2003my}, without invoking the $\delta N$-formalism. A recent study \cite{Suyama:2007bg} confirmed, in the same framework, the equivalence of results based on the $\delta N$-formalism. Hence, without further ado, we apply the $\delta N$-formalism to the case of instant modulated preheating.

According to (\ref{fnl}) with (\ref{partialNpartialxi}), the additional contribution to $f_{NL}$ due to $\xi$ is proportional to
\begin{eqnarray}
 N_{\xi\xi}=\frac{3}{2} \left(\gamma_1^2-\gamma_2\right)\,, \label{partNpartxx}
\end{eqnarray}
where we defined
\begin{eqnarray}
\gamma_n\equiv \frac{1}{h}\frac{\partial^n h}{\partial \xi^n}\bigg|_{t_i}\,.
\end{eqnarray}
 If we take a simple exponential dependence $h\sim e^{\beta \xi}$, this additional term vanishes identically. Further, it is easily checked that the same holds true for higher order correlation functions, such as those measured by the non linearity parameters $\tau_{NL}$ and $g_{NL}$ of \cite{Byrnes:2006vq} (see also \cite{Suyama:2007bg}). Thus, the most simple realization of the above mechanism is not expected to generate non-Gaussianities in addition to the intrinsic ones of the isocurvature field, which can still be quite large in ekpyrotic models \cite{Lehners}.This has to be contrasted with the negligble non-Gaussianities in single field inflationary models (see however \cite{Chen:2006nt} for more intricate single field models), the moderately large NG in the curvaton scenario \cite{Sasaki:2006kq} or after a reflection of fields on a sharp boundary in ekpyrotic models \cite{Lehners}, and the large NG due to a conversion during the ekpyrotic phase as examined in \cite{Creminelli:2007aq,Buchbinder:2007tw,Koyama:2007if}.

As an example of the latter, consider a transition from the unstable two field scaling solution (reviewed in Appendix \ref{ap1}) to the single field, old ekpyrotic attractor solution, as discussed in \cite{Koyama:2007if}; the curved field trajectory responsible for the conversion causes an evolution of perturbations even after horizon crossing. This evolution is the primary cause for the generation of non-Gaussianities. Again, using the $\delta N$-formalism, one can compute $f_{NL}$ analytically to \cite{Koyama:2007if}
\begin{eqnarray}
f_{NL}^{\mbox{\tiny \cite{Koyama:2007if}}}=\frac{5}{12}c_i^2\,,\label{fnlKoyama}
\end{eqnarray}
where $c_i$ is the exponent in the potential (\ref{potentialW}) of the field which becomes subdominant at late times \footnote{This result was also confirmed numerically in \cite{Koyama:2007if}. }. Since $c_i$ has to be reasonably large in the ekpyrotic phase in order for fast roll to occur, the non-linearity parameter becomes quite large too; indeed, current observations put the model already under pressure \cite{Koyama:2007if}. Accordingly, such a conversion before the bounce will be refuted if large non-Gaussianities are not found in the near future. 

A conversion due to a reflection of fields on a sharp boundary, during a phase where scalar kinetic energy dominates, leads to $f_{NL}\sim\mathcal{O}(c_i)$, smaller than (\ref{fnlKoyama}) but still potentially measurable in upcoming experiments \cite{Lehners}.
Our proposal with $h\sim e^{\beta \xi}$ falls into the same class. Of course, more complicated $h(\xi)$ could lead to potentially large non-Gaussianities even in our proposal. Assuming $N_\xi\gg N_\phi$, we can estimate the additional contribution to the non-linearity parameter to
\begin{eqnarray}
f_{NL}^{ad}\approx\frac{5}{9}\left(1-\frac{\gamma_2}{\gamma_1^2}\right)\,,
\end{eqnarray}
where we used (\ref{partNpartxx}) and (\ref{partialNpartialxi}) in (\ref{fnl}).
At this point a better understanding of $h(\xi)$ is needed in order to compute $\gamma_i$ and provide more concrete predictions. The same holds true for the amplitude of the power spectrum \cite{inpreparation}. In absence of such a study, which requires a much improved understanding of the underlying string theoretical construction, one might favor the simple exponential dependence put forward here with $f_{NL}^{ad}\approx 0$.

Before we conclude, we would like to mention the possibility to discern modulated preheating from standard single field inflation by testing the consistency relation between the amplitude of curvature perturbations, the amplitude of tensor modes and the tensor spectral index; in modulated reheating, this relation need not be the same as in simple inflationary models \cite{Matarrese:2003tk}. Again, a better understanding of the underlying stringy construction of the new ekpyrotic scenario is needed in order to make any concrete predictions.

\section{Conclusion}

 In brief, we presented a mechanism for imprinting scalar perturbations onto the matter content after the bounce in ekpyrotic models of the universe (or after an inflationary phase), providing a realization of modulated perturbations from instant preheating.  The spectrum is given by the one of a scalar isocurvature field, which determines the magnitude of the coupling constant between preheat matter and Fermions. The model is independent of the details of the bounce in ekpyrotic models, as well as the spectrum in the adiabatic mode before reheating. We estimate non-Gaussianities and find no extra contribution in the most simple realization of modulated instant preheating, in contrast to models relying on the curvaton mechanism in one way or another. This makes our proposal appealing, especially as long as primordial non-Gaussianities are not observed. \footnote{{\em Note added}: during the completion of this work a related article by Suyama and Yamaguchi appeared in \cite{Suyama:2007bg}, discussing non-Gaussianities in models of modulated reheating after inflation with multiple isocurvature fields. Their study is based on the $\delta N$-formalism, similar to the approach followed in this paper. The difference lies in the focus on inflation and the absence of instant preheating. Therefore, the imprint of fluctuation onto the matter content in \cite{Suyama:2007bg} occurs over many oscillations of the inflaton field during ordinary preheating, so that  the transition to a radiation dominated epoch is less sharp. In fact, this is the origin of the nontrivial $Q(\Gamma/H_0)$ in \cite{Suyama:2007bg}, which complicates the computations and yields potentially large NG. We agree with the approach taken in \cite{Suyama:2007bg} and it should be straightforward to generalize their results to ordinary preheating after ekpyrosis.}

\begin{acknowledgments}
We would like to thank Scott Watson for many discussions resulting in this study and initial collaboration, as well as  Diana Battefeld, Robert Brandenberger, Jean-Luc Lehners and Paul Steinhardt for comments on the draft. T.B. is supported in part by PPARC grant PP/D507366/1 and the Council on Science and Technology at Princeton University.
\end{acknowledgments}

\appendix

\section{Generation of Scale Invariant Isocurvature perturbations in a contracting Universe \label{ap1}}
For pedagogical reasons, we would like to review the origin of scale invariant perturbations in an isocurvature mode during a contracting phase  of the universe in the most simple cases, following closely \cite{Koyama:2007ag,Koyama:2007mg}.

\subsection{Background Dynamics}
Consider two scalar fields $\varphi_i$, $i=1,2$, with canonical kinetic terms and uncoupled, exponential potentials
\begin{eqnarray}
W=-\sum_{i=1}^{2}V_ie^{-c_i\varphi_i}\,, \label{potentialW}
\end{eqnarray}
with $c_i\equiv \sqrt{2/p_i}>0$ and $V_i>0$. Their equations of motion are
\begin{eqnarray}
\ddot{\varphi}_i+3H\dot{\varphi}_i=-\frac{d W}{d \varphi_i}
\end{eqnarray}
and the Friedman equations read
\begin{eqnarray}
3H^2&=&W+\sum_{i=1}^{2}\frac{1}{2}\dot{\varphi}_i^2\,,\\
\dot{H}&=&-\sum_{i=1}^{2}\frac{1}{2}\dot{\varphi}_i^2\,.
\end{eqnarray}
Here and in the following we set $8\pi G\equiv 1$. Given the potential in (\ref{potentialW}), there are two attractor solutions with either $\varphi_1$ or $\varphi_2$ dominating. Each attractor corresponds to a single field ekpyrotic solution. However, there is an additional, exact, multi-field scaling solution to the equations of motion, for which the ratio of the fields' kinetic and potential energy remains constant. This solution is easily obtained after a field redefinition into an \emph{adiabatic field}
\begin{eqnarray}
\phi\equiv \frac{c_2\varphi_1+c_1\varphi_2}{\sqrt{c_1^2+c_2^2}}
\end{eqnarray}
and an \emph{isocurvature field}
\begin{eqnarray}
\xi\equiv \frac{c_1\varphi_1-c_2\varphi_2}{\sqrt{c_1^2+c_2^2}}\,,
\end{eqnarray}
so that the potential becomes
\begin{eqnarray}
W=-U(\xi)e^{-c\phi}\,,
\end{eqnarray}
where $c^{-2}\equiv c_1^{-2}+c_2^{-2}$ and
\begin{eqnarray}
U(\xi)=V_1e^{-(c_1/c_2)c\xi}+V_2e^{(c_2/c_1)c\xi}\,.
\end{eqnarray}
One can  check that $-U$ has a maximum at
\begin{eqnarray}
\xi_0=\frac{1}{\sqrt{c_1^2+c_2^2}}\ln\left(\frac{c_1^2V_1}{c_2^2V_2}\right)\,.
\end{eqnarray}
The scaling solution corresponds to $\xi=\xi_0$, whereas $\phi$ rolls down the exponential potential \footnote{In case the fields move away from the scaling solution, one can still define instantaneous isocurvature and adiabatic fields. Since we assume the validity of the scaling solution up until the end of preheating, the initial decomposition remains valid  and we refer to $\phi$ and $\xi$ as the adiabatic and isocurvature field throughout. }. Obviously, this solution is unstable, and it is indeed the tachyonic instability which gives rise to a scale invariant spectrum of perturbations in $\xi$, given that the universe is contracting with an equation of state parameter satisfying $w\gg 1$. The presence of an instability is not a peculiarity of this particular potential, but a necessity \cite{Tolley:2007nq}. Naturally, such multi-field ekpyrotic proposals are rather contrived, as long as no compelling reason is found to start with initial conditions in close proximity to the repeller; one could postulate an additional pre-ekpyrotic phase \cite{Buchbinder:2007tw}, to drive $\xi$ towards the repeller at early times via an additional potential with positive mass squared and a coupling to Fermions \footnote{The coupling to Fermions is needed in order to dissipate energy while $\xi$ is oscillating around $\xi_0$, because blueshifting would amplify oscillations otherwise.}. Ignoring this problem, let us give the model the benefit of the doubt and assume that $\xi$ stays close to $\xi_0$ throughout, so that $\dot{\xi}\approx 0$ and $U(\xi)\approx U(\xi_0)=\mbox{const}$. It follows that the Friedmann equation yields power law contraction $a\propto (-t)^p$
with $p\equiv 2/c^2$. It will be useful to introduce
\begin{eqnarray}
\bar{\varepsilon}\equiv\frac{3}{2}(1+w)=\frac{1}{p}=\frac{1}{2\varepsilon}\gg 1\,,
\end{eqnarray}
where we used the fast roll parameter
\begin{eqnarray}
\varepsilon\equiv \left(\frac{W}{W_{,\phi}}\right)^2
\end{eqnarray}
in the last step (here $(\cdot)_{,\phi}$ denotes a partial derivative with respect to $\phi$). Further, the Klein Gordon equation for the adiabatic field yields $\dot{\phi}=\sqrt{2p}/t$. Afterwards, we will switch to conformal time $\eta=\int1/a\,dt$ so that  $a\propto (-\eta)^{1/(\bar{\varepsilon}-1)}$.

\subsection{Perturbations \label{ap2}}
A useful quantity is the Sasaki-Mukhanov variable, for multiple fields $\phi_I$ defined via
\begin{eqnarray}
Q_I=\delta \phi_I+\frac{\dot{\phi}_I}{H}\psi\,.
\end{eqnarray}
Here, $\psi$ is the Newtonian potential (the diagonal metric perturbation in the spatial directions). $Q_I$ has the easy interpretation of the $I$'th field perturbation in the spatially flat gauge ($\psi=0$). The comoving curvature perturbation $\mathcal{R}$ \footnote{On large scales, $\mathcal{R}$ is identical to $-\zeta$.} is related to the $Q_I$ via $\mathcal{R}=\sum_I\dot{\phi}_iQ_I/\sum_J\dot{\phi}_J^2$. The Fourier components of $Q_I$ satisfy (see e.g. \cite{Gordon:2000hv} for a derivation from first principles)
\begin{eqnarray}
0&=&\ddot{Q}_I+3H\dot{Q}_I+\frac{k^2}{a^2}Q_I\\
\nonumber &&+\sum_J\left(W_{,\phi_I\phi_J}-\frac{1}{a^3}\left(\frac{a^3}{H}\dot{\phi}_I\dot{\phi}_J\right)^{\!.}\right)Q_J\,.
\end{eqnarray}
It is already obvious at this stage that one should generally expect a coupling between the different $Q_I$'s. However, working in the $\psi=0$ gauge and decomposing perturbations into an adiabatic one along the field trajectory and an isocurvature one perpendicular to it, it becomes evident that the isocurvature mode seeds the adiabatic mode only if the trajectory in field space is curved \cite{Gordon:2000hv}. Since our scaling solution corresponds to a straight line in field space, no conversion of the isocurvature mode to the adiabatic one will occur and the equations decouple if they are written in terms of $Q_\phi=\delta\phi$ and $Q_\xi=\delta\xi$.

To be precise, using $\dot{\xi}=0$ as well as $W_{,\xi}=0$ we arrive at
\begin{eqnarray}
0&=&\ddot{\delta \xi}+3H\dot{\delta \xi}+\left(\frac{k^2}{a^2}+W_{,\xi\xi}\right)\delta\xi\,,\\
0&=&\ddot{\delta \phi}+3H\dot{\delta \phi}+\left(\frac{k^2}{a^2}+W_{,\phi\phi}-\frac{1}{a^3}\left(\frac{a^3}{H}\dot{\phi}^2\right)^\cdot\right)\delta\phi\,.
\end{eqnarray}
Introducing $u_\xi\equiv a\delta \xi$, $u_\phi\equiv a\delta \phi$ and working in conformal time, the above equations reduce after some algebra to
\begin{eqnarray}
0=u_I^{\prime\prime}+\left(k^2-\frac{\beta_I}{\eta^2}\right)u_I \label{eomui}
\end{eqnarray}
with $I=\chi,\phi$ and
\begin{eqnarray}
\beta_\xi&\equiv& \frac{2\bar{\varepsilon}^2-7\bar{\varepsilon}+2}{(\bar{\varepsilon}-1)^2}\,,\label{betaxi}\\
\beta_\phi&\equiv&\frac{2-\bar{\varepsilon}}{(\bar{\varepsilon}-1)^2}\,,\label{betaphi}
\end{eqnarray}
where we used the background solution of the previous section.

Imposing vacuum initial conditions well inside the horizon ($k^2\eta^2\gg\beta_I$ so that $\beta_I$ can be neglected) we have
\begin{eqnarray}
u_I=\frac{1}{\sqrt{2k}}\left(b({\bf k} )e^{-ik\eta}+b^{+}({\bf{k}})e^{ik\eta}\right)\,,
\label{soluini}
\end{eqnarray}
where $b$  satisfies $\langle b({\bf k})b^+(-\tilde{{\bf k}})\rangle=\delta^3({\bf k}-\tilde{{\bf k}})$ and $b|0\rangle=0$. If we keep the $\beta_I$ term in the equation of motion  we arrive at
\begin{eqnarray}
u_I=\sqrt{-k\eta}\left(C({\bf k})H_{\nu_I}^{(1)}(-k\eta)+C^+(-{\bf k})H_{\nu_I}^{(2)}(-k\eta)\right)\,,\label{nextuI}
\end{eqnarray}
where $H_{\nu_I}^{(1,2)}$ are Hankel functions of the first and second kind respectively, with index
\begin{eqnarray}
\nu_{I}=\frac{1}{2}\sqrt{1+4\beta_I}\,.\label{nui}
\end{eqnarray}
Expanding the Hankel functions for large arguments results in
\begin{eqnarray}
H_{\nu_I}^{(1,2)}\approx\sqrt{\frac{2}{\pi x}}e^{\pm i (x-(1+2\nu_I)\pi/2)}\,.
\end{eqnarray}
Here, the ``$+$'' corresponds to $H_{\nu_I}^{(1)}$ and the ``$-$'' to $H_{\nu_I}^{(2)}$. Using this expansion, we can match $u_I$ from (\ref{nextuI}) to (\ref{soluini}) and determine $C$ to
\begin{eqnarray}
C({\bf k})=\sqrt{\frac{\pi}{4k}} e^{\frac{i\pi}{4}(2\nu_I+1)} b({\bf k})\,.
\end{eqnarray}
After horizon crossing, that is in the limit $-k\eta\rightarrow 0$ but $-\ln(-k\eta)$ not too large, we can use the small argument limit of the Hankel functions
\begin{eqnarray}
H_{\nu_I}^{(1,2)}(x)\approx\pm \Gamma(\nu_I)\left(\frac{2}{x}\right)^{\nu_I}\frac{1}{i\pi}\,,
\end{eqnarray}
where $\Gamma$ is the Gamma function. Bringing everything together, we arrive at
\begin{eqnarray}
u_I=\frac{1}{\sqrt{4\pi k}}2^{\nu_I}\Gamma(\nu_I)(-k\eta)^{\frac{1}{2}-\nu_I}\tilde{b}({\bf k})
\end{eqnarray}
where
\begin{eqnarray}
\tilde{b}({\bf k})\equiv \frac{1}{i}\left(e^{\frac{i\pi}{4}(2\nu_I+1)}b({\bf k})-e^{-\frac{i\pi}{4}(2\nu_I+1)}b^+(-{\bf k})\right)
\end{eqnarray}
satisfies also $\langle \tilde{b}({\bf k})\tilde{b}^+(\tilde{{\bf k}})\rangle=\delta^3({\bf k}-\tilde{{\bf k}})$. Thus we have $\delta \phi=u_\phi/a\propto k^{-\nu_\phi}$ and $\delta \xi=u_\xi/a\propto k^{-\nu_\xi}$. Since the power spectrum is defined via
\begin{eqnarray}
\delta({\bf k}-{\bf k}^\prime)\frac{2\pi^2}{k^3}\mathcal{P}_I\equiv\frac{1}{a^2}\langle u_I({\bf k}) u^+_I({\bf k}^\prime)\rangle
\end{eqnarray}
we can easily read off the scalar spectral index to
\begin{eqnarray}
n_I-1&\equiv&\frac{d\ln \mathcal{P}_I}{d\ln k}\\
&=&3-2\nu_I\,.
\end{eqnarray}
 If we further expand $\nu_I$ from (\ref{nui}) with (\ref{betaxi}) and (\ref{betaphi}) in terms of the fast roll parameter $\varepsilon=1/2\bar{\varepsilon}\ll 1$ we arrive at
\begin{eqnarray}
n_\phi-1&\simeq&2+4\varepsilon\,,\\
n_\xi-1&\simeq&4 \varepsilon\,.
\end{eqnarray}

As expected, the adiabatic mode carries the deep blue spectrum from single field ekpyrosis, whereas the isocurvature mode has a nearly scale invariant spectrum with a small blue tilt. However, if slightly different potentials are considered, the index can easily acquire a red tilt \cite{Lehners:2007ac,Buchbinder:2007ad} \footnote{Note that $w\neq const$ for more complicated potentials, so that one has to take into account the time dependence of $\varepsilon$. Further, a second fast roll parameter related to the second derivative of the potential does not vanish any more and it is indeed this second parameter which can turn the blue tilt into a red one.}.

At this point we would like to emphasize again that the tachyonic instability of the scaling solution is crucial for obtaining a nearly scale invariant spectrum in a contracting universe; to see this, let us re-examine how $n_\xi-1\approx 0$ arose above: from $n_\xi-1=3-2\nu_\xi$ we see that $\nu_\xi\approx 3/2$ has to hold, which in turn requires $\beta_\xi$ to be close to $2$, via the definition of $\nu_\xi$ (\ref{nui}). If we now look back how $\beta_\xi$ arose in the equation of motion (\ref{eomui}), we see that the leading order contribution during fast roll ($\mathcal{H}$ and $\mathcal{H}^\prime$ are small) stems from $\beta_\xi \sim -W_{,\xi\xi}a^2\tau^2$. As a result, the second derivative of $W$ has to be negative, which in turn means that $\xi$ exhibits a tachyonic instability (recall that $W_{,\xi}=0$). This is true more generally \cite{Tolley:2007nq}, and not an artifact of the simple model employed here.


\begin{thebibliography}{99}

\bibitem{Khoury:2001wf}
  J.~Khoury, B.~A.~Ovrut, P.~J.~Steinhardt and N.~Turok,
 ``The ekpyrotic universe: Colliding branes and the origin of the hot big
  bang,''
  Phys.\ Rev.\  D {\bf 64}, 123522 (2001)
  [arXiv:hep-th/0103239].

\bibitem{Khoury:2001bz}
  J.~Khoury, B.~A.~Ovrut, N.~Seiberg, P.~J.~Steinhardt and N.~Turok,
  ``From big crunch to big bang,''
  Phys.\ Rev.\  D {\bf 65}, 086007 (2002)
  [arXiv:hep-th/0108187].

\bibitem{Steinhardt:2001st}
  P.~J.~Steinhardt and N.~Turok,
  ``Cosmic evolution in a cyclic universe,''
  Phys.\ Rev.\  D {\bf 65}, 126003 (2002)
  [arXiv:hep-th/0111098].


\bibitem{Linde:2007fr}
  A.~Linde,
  ``Inflationary Cosmology,''
  arXiv:0705.0164 [hep-th].

\bibitem{Brandenberger:2001bs}
  R.~Brandenberger and F.~Finelli,
  ``On the spectrum of fluctuations in an effective field theory of the
  ekpyrotic universe,''
  JHEP {\bf 0111}, 056 (2001)
  [arXiv:hep-th/0109004].

\bibitem{Durrer:2002jn}
R.~Durrer and F.~Vernizzi,
``Adiabatic perturbations in pre big bang models: Matching conditions and
scale invariance,''
Phys.\ Rev.\ D {\bf 66}, 083503 (2002)
[arXiv:hep-ph/0203275].


\bibitem{Creminelli:2004jg}
  P.~Creminelli, A.~Nicolis and M.~Zaldarriaga,
  ``Perturbations in bouncing cosmologies: Dynamical attractor vs scale
  invariance,''
  Phys.\ Rev.\  D {\bf 71}, 063505 (2005)
  [arXiv:hep-th/0411270].

\bibitem{Battefeld:2004mn}
  T.~J.~Battefeld, S.~P.~Patil and R.~Brandenberger,
  ``Perturbations in a bouncing brane model,''
  Phys.\ Rev.\  D {\bf 70}, 066006 (2004)
  [arXiv:hep-th/0401010].


\bibitem{Lehners:2007ac}
  J.~L.~Lehners, P.~McFadden, N.~Turok and P.~J.~Steinhardt,
  ``Generating ekpyrotic curvature perturbations before the big bang,''
  arXiv:hep-th/0702153.

\bibitem{Buchbinder:2007ad}
  E.~I.~Buchbinder, J.~Khoury and B.~A.~Ovrut,
  ``New ekpyrotic cosmology,''
  arXiv:hep-th/0702154.

\bibitem{Creminelli:2007aq}
  P.~Creminelli and L.~Senatore,
  ``A smooth bouncing cosmology with scale invariant spectrum,''
  arXiv:hep-th/0702165.

\bibitem{Buchbinder:2007tw}
  E.~I.~Buchbinder, J.~Khoury and B.~A.~Ovrut,
  ``On the Initial Conditions in New Ekpyrotic Cosmology,''
  arXiv:0706.3903 [hep-th].


\bibitem{Koyama:2007ag}
  K.~Koyama, S.~Mizuno and D.~Wands,
  ``Curvature perturbations from ekpyrotic collapse with multiple fields,''
  arXiv:0704.1152 [hep-th].

\bibitem{Koyama:2007mg}
  K.~Koyama and D.~Wands,
  ``Ekpyrotic collapse with multiple fields,''
  JCAP {\bf 0704}, 008 (2007)
  [arXiv:hep-th/0703040].

\bibitem{Koyama:2007if}
  K.~Koyama, S.~Mizuno, F.~Vernizzi and D.~Wands,
  ``Non-Gaussianities from ekpyrotic collapse with multiple fields,''
  arXiv:0708.4321 [hep-th].

\bibitem{Lehners:2006pu}
  J.~L.~Lehners, P.~McFadden and N.~Turok,
  ``Colliding branes in heterotic M-theory,''
  arXiv:hep-th/0611259.

\bibitem{Lehners:2006ir}
  J.~L.~Lehners, P.~McFadden and N.~Turok,
  ``Effective actions for heterotic M-theory,''
  arXiv:hep-th/0612026.

\bibitem{Lehners:2007nb}
  J.~L.~Lehners and N.~Turok,
  ``Bouncing Negative-Tension Branes,''
  arXiv:0708.0743 [hep-th].

\bibitem{Lehners}
J.~L.~Lehners and P.~J.~Steinhardt, in preparation.

\bibitem{Lyth:2001nq}
  D.~H.~Lyth and D.~Wands,
  ``Generating the curvature perturbation without an inflaton,''
  Phys.\ Lett.\  B {\bf 524}, 5 (2002)
  [arXiv:hep-ph/0110002].

\bibitem{Notari:2002yc}
  A.~Notari and A.~Riotto,
  ``Isocurvature perturbations in the ekpyrotic universe,''
  Nucl.\ Phys.\  B {\bf 644}, 371 (2002)
  [arXiv:hep-th/0205019].


\bibitem{Finelli:2007tr}
  F.~Finelli, P.~Peter and N.~Pinto-Neto,
  ``Spectra of primordial fluctuations in two-perfect-fluid regular bounces,''
  arXiv:0709.3074 [gr-qc].

\bibitem{Bozza:2005xs}
  V.~Bozza and G.~Veneziano,
  ``Regular two-component bouncing cosmologies and perturbations therein,''
  JCAP {\bf 0509}, 007 (2005)
  [arXiv:gr-qc/0506040].

\bibitem{Bozza:2005wn}
  V.~Bozza and G.~Veneziano,
  ``Scalar perturbations in regular two-component bouncing cosmologies,''
  Phys.\ Lett.\  B {\bf 625}, 177 (2005)
  [arXiv:hep-th/0502047].

\bibitem{Geshnizjani:2005hc}
  T.~J.~Battefeld and G.~Geshnizjani,
  ``A note on perturbations during a regular bounce,''
  Phys.\ Rev.\  D {\bf 73}, 048501 (2006)
  [arXiv:hep-th/0506139].

\bibitem{Battefeld:2005cj}
  T.~J.~Battefeld and G.~Geshnizjani,
  ``Perturbations in a regular bouncing universe,''
  Phys.\ Rev.\  D {\bf 73}, 064013 (2006)
  [arXiv:hep-th/0503160].


\bibitem{Kofman:2003nx}
  L.~Kofman,
  ``Probing string theory with modulated cosmological fluctuations,''
  arXiv:astro-ph/0303614.

\bibitem{Dvali:2003em}
  G.~Dvali, A.~Gruzinov and M.~Zaldarriaga,
  ``A new mechanism for generating density perturbations from inflation,''
  Phys.\ Rev.\  D {\bf 69}, 023505 (2004)
  [arXiv:astro-ph/0303591].

\bibitem{Bernardeau:2004zz}
  F.~Bernardeau, L.~Kofman and J.~P.~Uzan,
  ``Modulated fluctuations from hybrid inflation,''
  Phys.\ Rev.\ D {\bf 70}, 083004 (2004)
  [arXiv:astro-ph/0403315].


\bibitem{Enqvist:2003uk}
  K.~Enqvist, A.~Mazumdar and M.~Postma,
  ``Challenges in generating density perturbations from a fluctuating  inflaton
  coupling,''
  Phys.\ Rev.\  D {\bf 67}, 121303 (2003)
  [arXiv:astro-ph/0304187].

\bibitem{Tsujikawa:2003bn}
  S.~Tsujikawa,
  ``Cosmological density perturbations from perturbed couplings,''
  Phys.\ Rev.\  D {\bf 68}, 083510 (2003)
  [arXiv:astro-ph/0305569].


\bibitem{Matarrese:2003tk}
  S.~Matarrese and A.~Riotto,
  ``Large-scale curvature perturbations with spatial and time variations of
  the inflaton decay rate,''
  JCAP {\bf 0308}, 007 (2003)
  [arXiv:astro-ph/0306416].


\bibitem{Mazumdar:2003iy}
  A.~Mazumdar and M.~Postma,
  ``Evolution of primordial perturbations and a fluctuating decay rate,''
  Phys.\ Lett.\  B {\bf 573}, 5 (2003)
  [Erratum-ibid.\  B {\bf 585}, 295 (2004)]
  [arXiv:astro-ph/0306509].

\bibitem{Postma:2003jd}
  M.~Postma,
  ``Inhomogeneous reheating scenario with low scale inflation and/or MSSM  flat
  directions,''
  JCAP {\bf 0403}, 006 (2004)
  [arXiv:astro-ph/0311563].

\bibitem{Allahverdi:2004bk}
  R.~Allahverdi,
  ``Scenarios of modulated perturbations,''
  Phys.\ Rev.\  D {\bf 70}, 043507 (2004)
  [arXiv:astro-ph/0403351].

\bibitem{Ackerman:2004kw}
  L.~Ackerman, C.~W.~Bauer, M.~L.~Graesser and M.~B.~Wise,
  ``Light scalars and the generation of density perturbations during
  preheating or inflaton decay,''
  Phys.\ Lett.\  B {\bf 611}, 53 (2005)
  [arXiv:astro-ph/0412007].

\bibitem{Vernizzi:2005fx}
  F.~Vernizzi,
  ``Generating cosmological perturbations with mass variations,''
  Nucl.\ Phys.\ Proc.\ Suppl.\  {\bf 148}, 120 (2005)
  [arXiv:astro-ph/0503175].

\bibitem{Lyth:2005qk}
  D.~H.~Lyth,
  ``Generating the curvature perturbation at the end of inflation,''
  JCAP {\bf 0511}, 006 (2005)
  [arXiv:astro-ph/0510443].

\bibitem{Felder:1999pv}
  G.~N.~Felder, L.~Kofman and A.~D.~Linde,
   ``Inflation and preheating in NO models,''
  Phys.\ Rev.\ D {\bf 60}, 103505 (1999)
  [arXiv:hep-ph/9903350].

\bibitem{Sasaki:2006kq}
  M.~Sasaki, J.~Valiviita and D.~Wands,
  ``Non-gaussianity of the primordial perturbation in the curvaton model,''
  Phys.\ Rev.\  D {\bf 74}, 103003 (2006)
  [arXiv:astro-ph/0607627].


\bibitem{Tolley:2007nq}
  A.~J.~Tolley and D.~H.~Wesley,
  ``Scale-invariance in expanding and contracting universes from two-field
  models,''
  arXiv:hep-th/0703101.


\bibitem{Traschen:1990sw}
  J.~H.~Traschen and R.~H.~Brandenberger,
  ``PARTICLE PRODUCTION DURING OUT-OF-EQUILIBRIUM PHASE TRANSITIONS,''
  Phys.\ Rev.\  D {\bf 42}, 2491 (1990).

\bibitem{Shtanov:1994ce}
  Y.~Shtanov, J.~H.~Traschen and R.~H.~Brandenberger,
  ``Universe reheating after inflation,''
  Phys.\ Rev.\  D {\bf 51}, 5438 (1995)
  [arXiv:hep-ph/9407247].

\bibitem{Kofman:1994rk}
  L.~Kofman, A.~D.~Linde and A.~A.~Starobinsky,
  ``Reheating after inflation,''
  Phys.\ Rev.\ Lett.\  {\bf 73}, 3195 (1994)
  [arXiv:hep-th/9405187].

\bibitem{Kofman:1997yn}
  L.~Kofman, A.~D.~Linde and A.~A.~Starobinsky,
  ``Towards the theory of reheating after inflation,''
  Phys.\ Rev.\  D {\bf 56}, 3258 (1997)
  [arXiv:hep-ph/9704452].

\bibitem{Bassett:2005xm}
  B.~A.~Bassett, S.~Tsujikawa and D.~Wands,
  ``Inflation dynamics and reheating,''
  Rev.\ Mod.\ Phys.\  {\bf 78}, 537 (2006)
  [arXiv:astro-ph/0507632].



\bibitem{Kofman:2004yc}
  L.~Kofman, A.~Linde, X.~Liu, A.~Maloney, L.~McAllister and E.~Silverstein,
  ``Beauty is attractive: Moduli trapping at enhanced symmetry points,''
  JHEP {\bf 0405}, 030 (2004)
  [arXiv:hep-th/0403001].

\bibitem{Watson:2004aq}
  S.~Watson,
  ``Moduli stabilization with the string Higgs effect,''
  Phys.\ Rev.\  D {\bf 70}, 066005 (2004)
  [arXiv:hep-th/0404177].

\bibitem{Patil:2004zp}
  S.~P.~Patil and R.~Brandenberger,
  ``Radion stabilization by stringy effects in general relativity and  dilaton
  gravity,''
  Phys.\ Rev.\  D {\bf 71}, 103522 (2005)
  [arXiv:hep-th/0401037].

\bibitem{Battefeld:2005wv}
  T.~J.~Battefeld, S.~P.~Patil and R.~H.~Brandenberger,
  ``On the transfer of metric fluctuations when extra dimensions bounce or
  stabilize,''
  Phys.\ Rev.\  D {\bf 73}, 086002 (2006)
  [arXiv:hep-th/0509043].

\bibitem{Battefeld:2005av}
  T.~Battefeld and S.~Watson,
  ``String gas cosmology,''
  Rev.\ Mod.\ Phys.\  {\bf 78}, 435 (2006)
  [arXiv:hep-th/0510022].

\bibitem{Battefeld:2006cn}
  T.~Battefeld and N.~Shuhmaher,
  ``Predictions of dynamically emerging brane inflation models,''
  Phys.\ Rev.\  D {\bf 74}, 123501 (2006)
  [arXiv:hep-th/0607061].

\bibitem{Cremonini:2006sx}
  S.~Cremonini and S.~Watson,
  ``Dilaton dynamics from production of tensionless membranes,''
  Phys.\ Rev.\  D {\bf 73}, 086007 (2006)
  [arXiv:hep-th/0601082].

\bibitem{Greene:2007sa}
  B.~Greene, S.~Judes, J.~Levin, S.~Watson and A.~Weltman,
  ``Cosmological moduli dynamics,''
  arXiv:hep-th/0702220.

\bibitem{Erickson:2006wc}
  J.~K.~Erickson, S.~Gratton, P.~J.~Steinhardt and N.~Turok,
  ``Cosmic perturbations through the cyclic ages,''
  arXiv:hep-th/0607164.


\bibitem{Starobinski}
A.~A.~Starobinsky, JETP Lett. {\bf 42}, 152 (1985) [Pis. Hz. Esp.
Tor. Fizz. {\rm 42}, 124 (1985)].

\bibitem{Sasaki:1995aw}
  M.~Sasaki and E.~D.~Stewart,
  ``A General analytic formula for the spectral index of the density
  perturbations produced during inflation,''
  Prog.\ Theor.\ Phys.\  {\bf 95}, 71 (1996)
  [arXiv:astro-ph/9507001].

\bibitem{Nakamura:1996da}
  T.~T.~Nakamura and E.~D.~Stewart,
  ``The spectrum of cosmological perturbations produced by a multi-component
  inflaton to second order in the slow-roll approximation,''
  Phys.\ Lett.\  B {\bf 381}, 413 (1996)
  [arXiv:astro-ph/9604103].


\bibitem{Lyth:2004gb}
  D.~H.~Lyth, K.~A.~Malik and M.~Sasaki,
  ``A general proof of the conservation of the curvature perturbation,''
  JCAP {\bf 0505}, 004 (2005)
  [arXiv:astro-ph/0411220].

\bibitem{Lyth:2005fi}
  D.~H.~Lyth and Y.~Rodriguez,
  ``The inflationary prediction for primordial non-gaussianity,''
  Phys.\ Rev.\ Lett.\  {\bf 95}, 121302 (2005)
  [arXiv:astro-ph/0504045].

\bibitem{Seery:2005gb}
  D.~Seery and J.~E.~Lidsey,
  ``Primordial non-gaussianities from multiple-field inflation,''
  JCAP {\bf 0509}, 011 (2005)
  [arXiv:astro-ph/0506056].

\bibitem{Vernizzi:2006ve}
  F.~Vernizzi and D.~Wands,
  ``Non-Gaussianities in two-field inflation,''
  JCAP {\bf 0605}, 019 (2006)
  [arXiv:astro-ph/0603799].


\bibitem{Byrnes:2006vq}
  C.~T.~Byrnes, M.~Sasaki and D.~Wands,
  ``The primordial trispectrum from inflation,''
  Phys.\ Rev.\  D {\bf 74}, 123519 (2006)
  [arXiv:astro-ph/0611075].

\bibitem{Byrnes:2007tm}
  C.~T.~Byrnes, K.~Koyama, M.~Sasaki and D.~Wands,
  ``Diagrammatic approach to non-Gaussianity from inflation,''
  arXiv:0705.4096 [hep-th].

\bibitem{Battefeld:2006sz}
  T.~Battefeld and R.~Easther,
  ``Non-gaussianities in multi-field inflation,''
  JCAP {\bf 0703}, 020 (2007)
  [arXiv:astro-ph/0610296].


\bibitem{Battefeld:2007en}
  D.~Battefeld and T.~Battefeld,
  ``Non-Gaussianities in N-flation,''
  JCAP {\bf 0705}, 012 (2007)
  [arXiv:hep-th/0703012].

\bibitem{Lyth:2003im}
  D.~H.~Lyth and D.~Wands,
  ``Conserved cosmological perturbations,''
  Phys.\ Rev.\  D {\bf 68}, 103515 (2003)
  [arXiv:astro-ph/0306498].

\bibitem{Rigopoulos:2003ak}
  G.~I.~Rigopoulos and E.~P.~S.~Shellard,
  ``The separate universe approach and the evolution of nonlinear  superhorizon
  cosmological perturbations,''
  Phys.\ Rev.\  D {\bf 68}, 123518 (2003)
  [arXiv:astro-ph/0306620].



\bibitem{Israel:1966rt}
  W.~Israel,
  ``Singular hypersurfaces and thin shells in general relativity,''
  Nuovo Cim.\  B {\bf 44S10}, 1 (1966)
  [Erratum-ibid.\  B {\bf 48}, 463 (1967\ NUCIA,B44,1.1966)].

\bibitem{Deruelle:1995kd}
  N.~Deruelle and V.~F.~Mukhanov,
  ``On matching conditions for cosmological perturbations,''
  Phys.\ Rev.\  D {\bf 52}, 5549 (1995)
  [arXiv:gr-qc/9503050].

\bibitem{unpublished}
 S. Watson, unpublished work.

\bibitem{Suyama:2006rk}
  T.~Suyama and S.~Yokoyama,
  ``Generating the primordial curvature perturbations in preheating,''
  Class.\ Quant.\ Grav.\  {\bf 24}, 1615 (2007)
  [arXiv:astro-ph/0606228].

\bibitem{Byrnes:2005th}
  C.~T.~Byrnes and D.~Wands,
  ``Scale-invariant perturbations from chaotic inflation,''
  Phys.\ Rev.\  D {\bf 73}, 063509 (2006)
  [arXiv:astro-ph/0512195].

\bibitem{Spergel:2006hy}
  D.~N.~Spergel {\it et al.}  [WMAP Collaboration],
  ``Wilkinson Microwave Anisotropy Probe (WMAP) three year results:
  Implications for cosmology,''
  arXiv:astro-ph/0603449.

\bibitem{Creminelli:2005hu}
  P.~Creminelli, A.~Nicolis, L.~Senatore, M.~Tegmark and M.~Zaldarriaga,
  ``Limits on non-Gaussianities from WMAP data,''
  JCAP {\bf 0605}, 004 (2006)
  [arXiv:astro-ph/0509029].

\bibitem{Creminelli:2006rz}
  P.~Creminelli, L.~Senatore, M.~Zaldarriaga and M.~Tegmark,
  ``Limits on $f_NL$ parameters from WMAP 3yr data,''
  arXiv:astro-ph/0610600.

\bibitem{Hikage:2006fe}
  C.~Hikage, E.~Komatsu and T.~Matsubara,
  ``Primordial Non-Gaussianity and Analytical Formula for Minkowski Functionals
  of the Cosmic Microwave Background and Large-scale Structure,''
  Astrophys.\ J.\  {\bf 653}, 11 (2006)
  [arXiv:astro-ph/0607284].

\bibitem{Komatsu:2001rj}
  E.~Komatsu and D.~N.~Spergel,
  ``Acoustic signatures in the primary microwave background bispectrum,''
  Phys.\ Rev.\  D {\bf 63}, 063002 (2001)
  [arXiv:astro-ph/0005036].

\bibitem{Planck}
    [Planck Collaboration],
  ``Planck: The scientific programme,''
  arXiv:astro-ph/0604069; http://www.rssd.esa.int/index.php?project=Planck


\bibitem{Zaldarriaga:2003my}
  M.~Zaldarriaga,
  ``Non-Gaussianities in models with a varying inflaton decay rate,''
  Phys.\ Rev.\  D {\bf 69}, 043508 (2004)
  [arXiv:astro-ph/0306006].


\bibitem{Suyama:2007bg}
  T.~Suyama and M.~Yamaguchi,
  ``Non-Gaussianity in the modulated reheating scenario,''
  arXiv:0709.2545 [astro-ph].

\bibitem{Chen:2006nt}
  X.~Chen, M.~x.~Huang, S.~Kachru and G.~Shiu,
  ``Observational signatures and non-Gaussianities of general single field
  inflation,''
  JCAP {\bf 0701}, 002 (2007)
  [arXiv:hep-th/0605045].

\bibitem{inpreparation}
T.~Battefeld and S.~Watson, in preparation.

\bibitem{Gordon:2000hv}
  C.~Gordon, D.~Wands, B.~A.~Bassett and R.~Maartens,
  ``Adiabatic and entropy perturbations from inflation,''
  Phys.\ Rev.\  D {\bf 63}, 023506 (2001)
  [arXiv:astro-ph/0009131].



\end{thebibliography}
\end{document}